\documentclass[prb,aps,tightenlines,twocolumn,%superscriptaddress,
floatfix]{revtex4}
\usepackage{graphicx}
\usepackage{amssymb}
\usepackage{amsmath}
\usepackage{bm}
\usepackage{colordvi}
\usepackage{mathrsfs}
\def\up{\uparrow}
\def\down{\downarrow}

\begin{document}
\title{Comment on ``Laser-assisted spin-polarized transport in graphene tunnel
  junctions''} 
\author{Y. Zhou}
\author{M. W. Wu}
\thanks{Author to whom correspondence should be addressed}
\email{mwwu@ustc.edu.cn.}
\affiliation{Hefei National Laboratory for Physical Sciences at
  Microscale and Department of Physics, University of Science and
  Technology of China, Hefei, Anhui, 230026, China}

\date{\today}

\maketitle

Recently, Ding {\em et al.}\cite{Berakdar_12} investigated spin-polarized
transport in graphene irradiated by a linearly polarized laser field.
There are several serious problems in their model, such as the violation of the
charge conservation in the graphene region, the incorrect application of the 
rotating-wave approximation (RWA) and even the wrong Green's functions. 
In the following, we discuss these problems point by point.

We first demonstrate that their approach violates the charge 
conservation in the graphene region. 
Our starting point is Eq.~(24) in Ref.~\onlinecite{Berakdar_12},
which gives the time-averaged current flowing into the left lead. 
That equation reads
\begin{eqnarray}
  \langle I_L \rangle&=&-\frac{e}{\hbar} \sum_{\tau n} \int \frac{d\varepsilon}{2\pi} \;
  {\rm Tr} \left\{\left[{G}^{\tau\tau,r}_{0n}(\varepsilon)
      -{G}^{\tau\tau,a}_{0n}(\varepsilon) \right]
    \Sigma_{L,n0}^{\tau,<}(\varepsilon) \right. \nonumber \\ 
  && \hspace{-0.35cm} 
    {}+ \sum_{\alpha n_1 n_2} G^{\tau\tau,r}_{0n_1}(\varepsilon)
    \Sigma_{\alpha,n_1 n_2}^{\tau,<}(\varepsilon)
    G^{\tau\tau,a}_{n_2 n}(\varepsilon)
    \Sigma_{L,n0}^{\tau,a}(\varepsilon) \bigg\}
    \label{I_original}
\end{eqnarray}
where $\tau=\up(\down)$ represents the spin-up (-down) band; $\alpha=L~(R)$
stands for the left (right) lead; the chemical potential in the leads are
$\mu_{L,R}=\pm eV/2$. 
For convenience, we transform all relevant Green's functions and self-energies 
from the rotating reference frame into the fixed reference frame, i.e., 
\begin{eqnarray}
  \tilde{G}^{\tau,r}_{ss'}(t,t')=G^{\tau\tau,r}_{ss'}(t,t') 
  e^{-i\frac{\omega_0}{2}(st-s't')}, 
\end{eqnarray}
with $s=1~(-1)$ standing for the conduction (valence) band. 
Equation~(\ref{I_original}) is then rewritten as
\begin{eqnarray}
  \langle I_L \rangle&=&-\frac{e}{h} \sum_{\tau} \int {d\varepsilon}\;
  {\rm Tr}\left\{\left[ \tilde{G}^{\tau,r}_{00}(\varepsilon)
      -\tilde{G}^{\tau,a}_{00}(\varepsilon) \right]
    \tilde{\Sigma}_{L}^{\tau,a}f_L(\varepsilon) \right. \nonumber \\ 
  && {}+ \sum_{n,\alpha} \tilde{G}^{\tau,a}_{0n}(\varepsilon)
    \tilde{\Sigma}_{L}^{\tau,a} \tilde{G}^{\tau,r}_{n0}(\varepsilon)
    \tilde{\Sigma}_{\alpha}^{\tau,a} f_\alpha(\varepsilon)
  \bigg\},
  \label{I_transform}
\end{eqnarray}
in which
\begin{eqnarray}
  && \hspace{-0.3cm}
  \tilde{G}^{\tau,r}_{nm}(\varepsilon)=
  \tilde{g}^{\tau,r}_{nm}(\varepsilon) + \frac{1}{2}\sum_{n_1 \alpha} 
  \tilde{g}^{\tau,r}_{n n_1}(\varepsilon) \tilde{\Sigma}^{\tau,r}_{\alpha} 
  \tilde{G}^{\tau,r}_{n_1 m}(\varepsilon), \label{G_big}\\
  && \hspace{-0.6cm}
  \tilde{g}^{\tau,r}_{nm,ss'}(\varepsilon)=\left\{ 
    \begin{array}{ll} \displaystyle 
      g^{\tau\tau,r}_{ss'}(\varepsilon+(n-s)\frac{\omega_0}{2}) \delta_{n,m} 
      & s=s' \\[0.2cm] \displaystyle 
      g^{\tau\tau,r}_{ss'}(\varepsilon+(n-s)\frac{\omega_0}{2}) \delta_{n,m+2s} 
      & s\ne s' \end{array}\right.\hspace{-0.2cm},\\
  && \hspace{0.25cm}
  \tilde{\Sigma}_{\alpha}^{\tau,a}=-\tilde{\Sigma}_{\alpha}^{\tau,r}
  =i{\Gamma}_{\alpha}^{\tau}
  \begin{pmatrix} 1 & 1 \\ 1 & 1 \end{pmatrix},
\end{eqnarray}
with $\Gamma_L^{\up\down}=\Gamma_R^{\up\down}=\Gamma_0(1\pm p)$.
The Green's functions $g^{\tau\tau,r}_{ss'}(\varepsilon)$
are given as Eqs.~(A.3)-(A.8) by Ding {\it et al.} in
Ref.~\onlinecite{Berakdar_12}. 
However, their expressions are mathematically incorrect, 
particularly the terms given by Eq.~(A.8) are even dimensionally 
incorrect. We present the correct formulae of
$g^{\tau\tau,r}_{ss'}(\varepsilon)$ in Appendix~\ref{correct_formulae}
in order to avoid misleading.

\begin{figure}[bp]
  \begin{center}
    \includegraphics[width=7.cm]{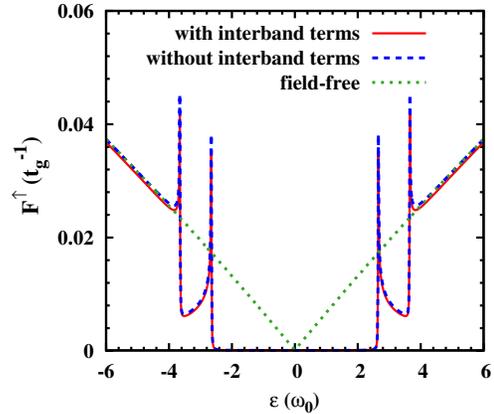}
  \end{center}
  \caption{ (Color online) $F^{\up}$ defined by Eq.~(\ref{f_spe}) as function of
    energy $\varepsilon$ with and without the interband term of
    $\tilde{g}^{\tau\tau,r}_{nm}(\varepsilon)$ in
    the case with the field strength $E_0=1200$~kV/cm and frequency
    $\omega_0=0.04 t_g$. The other parameters are $p=0.4$, $\Gamma_0=0.05 t_g$,
    $V_g=0$ and $D=3t_g$, same as those used in Fig.~1 in
    Ref.~\onlinecite{Berakdar_12}. We also plot the results in the
    field-free case as the green dotted curve.}
  \label{fig_f_spe} 
\end{figure}

Exchanging L and R in Eq.~(\ref{I_transform}),
one obtains the current flowing into the right lead,
\begin{eqnarray}
  \langle I_R \rangle&=&-\frac{e}{h} \sum_{\tau} \int {d\varepsilon} \;
  {\rm Tr}\left\{\left[ \tilde{G}^{\tau,r}_{00}(\varepsilon)
      -\tilde{G}^{\tau,a}_{00}(\varepsilon) \right]
    \tilde{\Sigma}_{R}^{\tau,a}f_R(\varepsilon) \right. \nonumber \\ 
  &&\left. {}+ \sum_{\alpha n} \tilde{G}^{\tau,a}_{0n}(\varepsilon)
    \tilde{\Sigma}_{R}^{\tau,a} \tilde{G}^{\tau,r}_{n0}(\varepsilon)
    \tilde{\Sigma}_{\alpha}^{\tau,a} f_\alpha(\varepsilon)
  \right\}.
\end{eqnarray}
Thus
\begin{eqnarray}
  && \hspace{-0.8cm}
 e\frac{dN_G}{dt}= -\langle I_L \rangle - \langle I_R \rangle 
  = \frac{e}{h} \sum_{\tau\alpha} \int {d\varepsilon} \;
  F^{\tau}(\varepsilon) {\Gamma}_{\alpha}^{\tau}f_{\alpha}(\varepsilon),\\
  && \hspace{-0.7cm}
  F^{\tau}(\varepsilon)= i[\overline{G}^{\tau,r}_{00}(\varepsilon)
  -\overline{G}^{\tau,a}_{00}(\varepsilon)]
  - \sum_{\alpha n}\overline{G}^{\tau,a}_{0n}(\varepsilon) 
  {\Gamma}_{\alpha}^{\tau} \overline{G}^{\tau,r}_{n0}(\varepsilon)
  \label{f_spe}
\end{eqnarray}
with $\overline{G}^{\tau,r}_{nm}(\varepsilon)=
\sum\limits_{ss'}\tilde{G}^{\tau,r}_{nm,ss'}(\varepsilon)$.
To show $F^{\tau}(\varepsilon)$ analytically, we first neglect
the interband term (i.e., $s\ne s'$) of 
$\tilde{g}^{\tau,r}_{nm}(\varepsilon)$
and obtain the approximate formula 
\begin{equation}
  F^{\tau}(\varepsilon)=\frac{2\pi D^{\tau}(\varepsilon)}
  {\Big|1+i\sum\limits_{\alpha s}
    {\Gamma}_{\alpha}^{\tau} \tilde{g}^{\tau,r}_{00,ss}(\varepsilon)\Big|^2}
\end{equation}
with $D^{\tau}(\varepsilon)=-\frac{1}{\pi}{\rm Im} \sum\limits_{s} 
\tilde{g}^{\tau,r}_{00,ss}(\varepsilon)$ being the
density of states of the isolated graphene under the laser field. 
Since $D^{\tau}(\varepsilon)$ is positive semidefinite,
$F^{\tau}(\varepsilon)$ is also positive semidefinite. 
Also considering $f_{\alpha}(\varepsilon)>0$, one obtains
${dN_G}/{dt}>0$, indicating the accumulation of the charge
in the graphene region.

To show the influence of the interband term, we plot the numerical results of 
$F^{\up}(\varepsilon)$ with and without the interband term
of $\tilde{g}^{\tau,r}_{nm}(\varepsilon)$ in Fig.~\ref{fig_f_spe}.\cite{F_down} 
% obtained via Eq.~(\ref{G_big})-(\ref{g_small_1}) and
% (\ref{gsmall_11})-(\ref{gsmall_12}), 
Here we use the field strength $E_0=1200$~kV/cm and frequency $\omega_0=0.04
t_g$.\cite{parameter} The other parameters are $p=0.4$, $\Gamma_0=0.05 t_g$,
$V_g=0$ and $D=3t_g$, same as those used in Fig.~1 in
Ref.~\onlinecite{Berakdar_12}. 
From this figure, one observes that $F^{\up}(\varepsilon)$ with and without the
interband term of $\tilde{g}^{\tau,r}_{nm}(\varepsilon)$ almost coincide and are
both positive semidefinite.
One clearly concludes that the approach reported by
Ding {\it et al.} {\it violates the charge
conservation in the graphene region}. 
Even worse, this problem not only appears in the laser-applied case discussed in
Ref.~\onlinecite{Berakdar_12}, but also appears in the field-free
case\cite{Berakdar_09,Berakdar_10} [$F^{\up}(\varepsilon)$ in that case are
plotted as green dotted curve in Fig.~\ref{fig_f_spe}(a)] and the case with the 
time-alternating gate voltage.\cite{Berakdar_11} 
Therefore, all results in these works are scientifically 
incorrect.\cite{nonequal} %!!!

\begin{figure}[t]
  \begin{center}
    \includegraphics[width=8.cm]{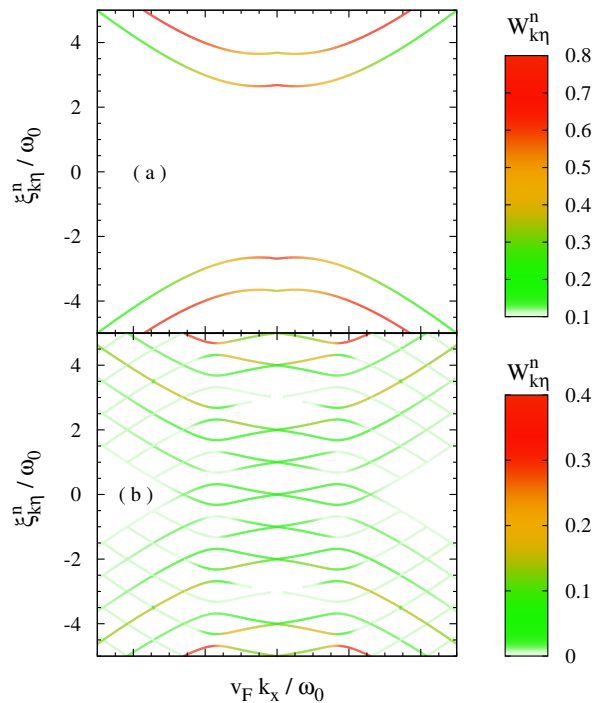}
  \end{center}
  \caption{ (Color online)  Quasi-energies of the sidebands 
    ${\xi}_{{\bf k}\eta}^{n}$ against the normalized momentum with (a) and 
    without (b) the RWA. The color coding represents the weight 
    $W_{{\bf k}\eta}^n$ of the corresponding sideband (note that it is a
    dimensionless quantity). 
    The parameters are the same as those of Fig.~\ref{fig_f_spe}. 
  }
  \label{fig_energy} 
\end{figure}

Another severe problem in Ref.~\onlinecite{Berakdar_12} is that they
mistakenly applied the RWA to the case
with strong laser field in the whole momentum regime. 
As shown in our recent work,\cite{Zhou_THz} the RWA is only valid for
the weak laser field at the momentum around the resonant point, i.e., 
$2v_{\rm F}k=\omega_0$ with $\omega_0$ being the frequency of the laser field. 
In order to make this issue more pronounced, we plot the sideband quasi-energies
and weights [defined by Eqs.~(10) and (11) in Ref.~\onlinecite{Zhou_THz}] with
and without the RWA in Figs.~\ref{fig_energy}(a) and (b), respectively.
The parameters are the same as those of Fig.~\ref{fig_f_spe}. 
Here we only show the case with the momentum along the current direction, 
i.e., the direction along the $x$ axis, as Ding {\em et al.}.\cite{Berakdar_12}
The Hamiltonian with and without the RWA are given by Eqs.~(10) and (7),
respectively in Ref.~\onlinecite{Berakdar_12}.
The corresponding eigenstates in these two cases can be obtained via the 
standard Floquet-Fourier approach widely used in the
literature.\cite{Zhou_THz,Shirley_65,Hanggi_rev,Oka_cur,Syzranov_08}
Comparing Figs.~\ref{fig_energy}(a) and (b), one finds that the quasi-energy
spectrum under the RWA is {\em qualitatively} different from the exact
one.
In particular, a huge gap opens around the Dirac point in the quasi-energy
spectrum under the RWA, in consistence with the gap in the bias dependence of
the differential conductance in Ref.~\onlinecite{Berakdar_12}. 
However, this gap is absent in the exact quasi-energy spectrum, as 
reported in the previous investigations on graphene under a 
linearly polarized laser.\cite{Zhou_THz,Syzranov_08,Calvo_APL} 

\begin{figure}[tbp]
  \begin{center}
    \includegraphics[width=7.cm]{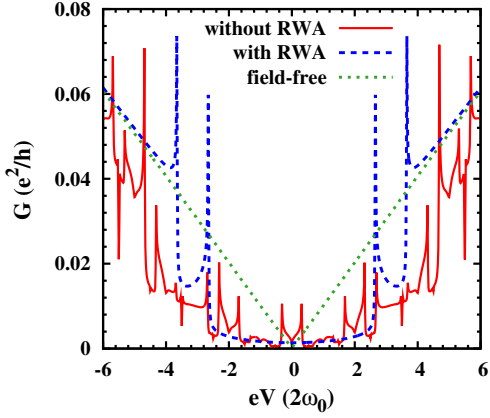}
  \end{center}
  \caption{ (Color online) The differential conductance as function of bias
    with and without the RWA. The parameters are the same as those of
    Fig.~\ref{fig_f_spe}. We also plot the results in the
    field-free case as the green dotted curve.
  }
  \label{fig_conduct} 
\end{figure}

Although the pronounced discrepancy in the quasi-energy spectrum with and
without the RWA is a convincing evidence of the invalidity of the
RWA in their cases, in order to nail down this issue, 
we further demonstrate that even under their framework, the differential
conductance with and without the RWA are {\em qualitatively} different.
Our approach is as follows. By exploiting the eigenstates
obtained above, which have the form
$|\Phi_{{\bf k}\eta}(t)\rangle=e^{-i\epsilon_{{\bf k}\eta}t}
  \sum\limits_{n=-\infty}^\infty e^{in\Omega t}|\phi^{n}_{{\bf k}\eta}\rangle$,
one obtains the Green's function of graphene without connecting
the leads,\cite{summation}
\begin{eqnarray}
  && \hspace{-0.3cm}
  \tilde{g}^{\tau,r}(t,t')= -i\theta(t-t') \sum_{{\bf k}\eta} 
  |\Phi_{{\bf k}\eta}(t)\rangle \langle \Phi_{{\bf k}\eta}(t')|, \\
  && \hspace{-0.8cm}
  \tilde{g}^{\tau,r}_{0N}(\varepsilon) = \left\{ 
    \begin{array}{ll} \displaystyle \sum\limits_{{\bf k} \eta n}
      \frac{|\phi^{n}_{{\bf k}\eta}\rangle \langle{\phi^{n-N/2}_{{\bf k}\eta}}|}
      {\varepsilon-\epsilon_{{\bf k}\eta}+n\omega_0+i0^+} & 
      \mbox{      $N$ is even} \\ 0 & \mbox{     $N$ is odd}
    \end{array}\right..
  \label{gsmall_noRWA}
\end{eqnarray}
We then  calculate the Green's function of graphene connected with the
leads via Eq.~(\ref{G_big}).
One obtains the current via Eq.~(\ref{I_transform}) and the differential 
conductance $G=-d\langle I_L \rangle/dV$.
We have examined that this approach can recover the
results under the RWA from the approach of
Ding {\it et al.} after correcting the errors in their Green's functions
[Eqs.~(A.3)-(A.8)] in that paper (see the correct formulae in
Appendix~\ref{correct_formulae}).\cite{examine} 
The conductances with and without the RWA are plotted in Fig.~\ref{fig_conduct}.
It is seen very clearly that the conductances in these two cases are 
{\em qualitatively} different, especially the pronounced 
gap around the Dirac point does not appear in the conductance without the RWA,
in consistence with the behaviour of the quasi-energy spectrum. 
Beyond all doubt, the above results can be seen as a {\em smoking gun} of the 
invalidity of the RWA in the cases discussed in
Ref.~\onlinecite{Berakdar_12}. Therefore, their main results, especially 
the pronounced conductance gap around the Dirac point, are scientifically
incorrect.

This work was supported by the National Basic Research Program of China under
Grant No.\ 2012CB922002 and the Strategic Priority Research Program of the
Chinese Academy of Sciences under Grant No. XDB01000000. 

%\appendix*
\begin{appendix}
\section{Green's functions under RWA}
\label{correct_formulae}
In Ref.~\onlinecite{Berakdar_12}, the Green's functions of graphene without
connecting the leads $g^{\tau\tau,r}_{s,s'}(\varepsilon)$
given by Eqs.~(A.3)-(A.8) are not the correct solution of Eq.~(A.2) in that
paper. In the following, we present the correct formulae of 
$g^{\tau\tau,r}_{s,s'}(\varepsilon)$ in the cases with $V_g=0$.
\begin{eqnarray}
  g^{\tau\tau,r}_{11}(\varepsilon)&=&-\frac{S}{2\pi v_{\rm F}^2}\bigg\{ \frac{1}{2}
  (\varepsilon+v_0)u_1 - \frac{\Delta^2-\varepsilon(\varepsilon+v_0)}
  {\sqrt{|\Delta^2-{\varepsilon}^2|}}
  \nonumber \\ &&
  {} \times \Big[ u_2\theta(\Delta-|\varepsilon|) 
  -u_3\theta(|\varepsilon|-\Delta) \Big] + D \bigg\} 
  \nonumber \\ && {}-\frac{iS\ {\rm sgn}(\varepsilon)}
  {4v_{\rm F}^2\sqrt{\varepsilon^2-\Delta^2}} \bigg[
  \theta\bigg(\Delta<|\varepsilon|<\sqrt{\frac{\omega_0^2}{4}+\Delta^2}\bigg)
  \nonumber \\ && {}\times
  \Big(\varepsilon-\sqrt{\varepsilon^2-\Delta^2}\Big)v_{-} 
  + \Big(\varepsilon+\sqrt{\varepsilon^2-\Delta^2}\Big)v_{+} 
  \nonumber \\ && {}\times
  \theta\bigg(\Delta<|\varepsilon|<\sqrt{(D-\frac{\omega_0}{2})^2+\Delta^2}
  \bigg) \bigg],
\label{gsmall_11}
\end{eqnarray}
\begin{eqnarray}
  g^{\tau\tau,r}_{22}(\varepsilon)&=&-\frac{S}{2\pi v_{\rm F}^2}\bigg\{ \frac{1}{2}
  (\varepsilon-v_0)u_1 + \frac{\Delta^2-\varepsilon(\varepsilon-v_0)}
  {\sqrt{|\Delta^2-{\varepsilon}^2|}}
  \nonumber \\ &&
  {} \times \Big[ u_2\theta(\Delta-|\varepsilon|) 
  -u_3\theta(|\varepsilon|-\Delta) \Big] - D \bigg\} 
  \nonumber \\ && {}-\frac{iS\ {\rm sgn}(\varepsilon)}
  {4v_{\rm F}^2\sqrt{\varepsilon^2-\Delta^2}} \bigg[
  \theta\bigg(\Delta<|\varepsilon|<\sqrt{\frac{\omega_0^2}{4}+\Delta^2}\bigg)
  \nonumber \\ && {}\times
  \Big(\varepsilon+\sqrt{\varepsilon^2-\Delta^2}\Big)v_{-} 
  + \Big(\varepsilon-\sqrt{\varepsilon^2-\Delta^2}\Big)v_{+} 
  \nonumber \\ && {}\times
  \theta\bigg(\Delta<|\varepsilon|<\sqrt{(D-\frac{\omega_0}{2})^2+\Delta^2}
  \bigg) \bigg],
\label{gsmall_22}
\end{eqnarray}
\begin{eqnarray}
  g^{\tau\tau,r}_{12}(\varepsilon)&=&g^{\tau\tau,r}_{21}(\varepsilon)
  =-\frac{S\Delta}{4\pi v_{\rm F}^2}
  \bigg\{ \frac{\omega_0}{\sqrt{|\Delta^2-{\varepsilon}^2|}} 
  \Big[ u_2\theta(\Delta-|\varepsilon|) 
  \nonumber \\ && {}
  -u_3\theta(|\varepsilon|-\Delta) \Big] + u_1 \bigg\} 
   -\frac{iS\ {\rm sgn}(\varepsilon)\Delta}
  {4v_{\rm F}^2\sqrt{\varepsilon^2-\Delta^2}} 
  \nonumber \\ && {} \times \Bigg[ v_{+}\ \theta\left(\Delta<|\varepsilon|
    <\sqrt{(D-\frac{\omega_0}{2})^2+\Delta^2} \right)
  \nonumber \\ && {} + v_{-}\ \theta\left(\Delta<|\varepsilon|<
      \sqrt{\frac{\omega_0^2}{4}+\Delta^2}\right) \Bigg],
  \label{gsmall_12}
\end{eqnarray}
with $S$ representing the area of sample.
$v_{\nu}=\frac{\omega_0}{2}+\nu\sqrt{|\varepsilon^2-\Delta^2|}$
with $\nu=0,\ \pm$ and
\begin{eqnarray}
  && \hspace{-1.cm}
  u_1={\rm ln}\left|\frac{(D-\frac{\omega_0}{2})^2
      +\Delta^2-{\varepsilon}^2}
    {\frac{\omega_0^2}{4}+\Delta^2-{\varepsilon}^2}\right|, \\
  && \hspace{-1.cm}
  u_2={\rm arctan}\frac{2D-\omega_0}
  {2\sqrt{\Delta^2-{\varepsilon}^2}}
  + {\rm arctan}\frac{\omega_0}
  {2\sqrt{\Delta^2-{\varepsilon}^2}}, \\
  && \hspace{-1.cm}
  u_3=\frac{1}{2}{\rm ln}\left|\frac
     {(D-\frac{\omega_0}{2}+\sqrt{\varepsilon^2-\Delta^2})
       (\frac{\omega_0}{2}+\sqrt{\varepsilon^2-\Delta^2})}
     {(D-\frac{\omega_0}{2}-\sqrt{\varepsilon^2-\Delta^2})
       (\frac{\omega_0}{2}-\sqrt{\varepsilon^2-\Delta^2})} \right|.
\end{eqnarray}

Comparing the above formulae with Eqs.~(A.3)-(A.8) in
Ref.~\onlinecite{Berakdar_12}, one observes
a lot of differences between them.
Specifically, in $g^{\tau\tau,r}_{11}$ and $g^{\tau\tau,r}_{22}$, they missed
$\omega_0/2$ in all $v_\nu$ terms.
The errors in the interband term are even more serious: their formula can be
written as
\begin{eqnarray}
  g^{\tau\tau,r}_{12}(\varepsilon)&=&g^{\tau\tau,r}_{21}(\varepsilon)
  =-\frac{S\Delta}{4\pi v_{\rm F}^2}
  \bigg[ u_1 \theta(\Delta-|\varepsilon|)
  \nonumber \\ && {}%\times
 +\frac{u_3}{\sqrt{|\Delta^2-{\varepsilon}^2|}} 
  \theta(|\varepsilon|-\Delta) \Big]  
   -\frac{iS\ {\rm sgn}(\varepsilon)\Delta}
  {4v_{\rm F}^2\sqrt{\varepsilon^2-\Delta^2}} 
  \nonumber \\ && {} \times
  \Bigg[\theta\bigg( \Delta<|\varepsilon|
  <\sqrt{(D-\frac{\omega_0}{2})^2+\Delta^2} \bigg)
  \nonumber \\ && {}
  - \theta\Bigg( \Delta<|\varepsilon|
  <\sqrt{\frac{\omega_0^2}{4}+\Delta^2} \Bigg)\Bigg].
  \label{gsmall_wrong}
\end{eqnarray}
It is seen that all terms in the above equation for
$|\varepsilon|>\Delta$ take the incorrect dimension.
Since there are so many errors in their formulae, 
these errors are not likely to only come from typos.

In order to show their errors more clearly, we calculate the DOS of the isolated
graphene in the field-free case, which has the form
\begin{eqnarray}
  && \hspace{-0.8cm}
  D^{\tau}_0(\varepsilon)= -\frac{1}{\pi}{\rm Im} 
  \sum_{s}\tilde{g}^{\tau,r}_{00,ss}(\varepsilon)\bigg|_{\Delta=0}
  \nonumber \\
  && \hspace{0.2cm} = -\frac{1}{\pi}{\rm Im} 
  \sum_{s}{g}^{\tau,r}_{00,ss}\Big(\varepsilon-s\frac{\omega_0}{2}\Big)\bigg|_{\Delta=0}.
\end{eqnarray}
From our formulae, one obtains
\begin{equation}
  D^{\tau}_0(\varepsilon)=\frac{S}{2\pi v_{\rm F}^2}\;|\varepsilon|\;
  \theta(|\varepsilon|<D),
\end{equation}
which is exactly the well-known formula 
of the DOS in graphene\cite{Neto_rev_09}
and is linear with energy. In contrast, from the formulae by Ding {\em et al.},
 one obtains
\begin{equation}
  D^{\tau}_0(\varepsilon)=\frac{S}{2\pi v_{\rm F}^2}
  \Big(|\varepsilon|-\frac{\omega_0}{2}\Big)
  \theta(|\varepsilon|<D).
\end{equation}
Obviously, the above formula is incorrect, especially it gives the negative DOS when 
$|\varepsilon|<\frac{\omega_0}{2}$.

\end{appendix}


\begin{thebibliography}{0}
\bibitem{Berakdar_12} K.-H. Ding, Z.-G. Zhu, and J. Berakdar, J. Phys.:
  Condens. Matter {\bf 24}, 266003 (2012).
\bibitem{F_down} The behaviour of $F^{\down}$ is similar to that of $F^{\up}$,
  and hence is not repeated here.  
\bibitem{parameter} In fact, the field strength $E_0$ given in
 Ref.~\onlinecite{Berakdar_12} is only one thousandth of the one used here.
 That number should be a typo as for such $E_0$, the dimensionless 
 quantity $\beta=ev_{\rm F}E_0/\omega_0^2=0.006$. 
 Such a small $\beta$ indicates that the laser field is too weak to influence
 the electric and transport properties of the system.\cite{Zhou_THz} 
\bibitem{Berakdar_09} K.-H. Ding, Z.-G. Zhu, and J. Berakdar, Phys. Rev. B 
  {\bf 79}, 045405 (2009).
\bibitem{Berakdar_10} K.-H. Ding, Z.-G. Zhu, Z.-H. Zhang, and J. Berakdar,
  Phys. Rev. B {\bf 82}, 155143 (2010). 
\bibitem{Berakdar_11} K.-H. Ding, Z.-G. Zhu, and J. Berakdar, Phys. Rev. B 
  {\bf 82}, 115433 (2011).
\bibitem{nonequal} It is noted that 
Eq.~(13) in Ref.~\onlinecite{Berakdar_10} obeys
 the charge conservation in graphene region. However, 
that equation even cannot be derived from their starting 
point Eq.~(11) in that paper, since 
  ${G}^{r}_{a}(\varepsilon)-{G}^{a}_{a}(\varepsilon) \ne
  {G}^{r}_{a}(\varepsilon) ({\Sigma}^{r}-{\Sigma}^{a})
  {G}^{a}_{a}(\varepsilon)$ under their framework 
(the definitions of all these symbols can
  be found in that paper). In fact, Eq.~(11) in that paper is equivalent to
the field-free form of Eq.~(\ref{I_transform}) in this comment and hence
 indeed violates the charge conservation as addressed in the main text.
  Refs.~\onlinecite{Berakdar_09} and \onlinecite{Berakdar_11} are in the 
  similar situation.
\bibitem{Zhou_THz} Y. Zhou and M. W. Wu, Phys. Rev. B {\bf 83}, 245436 (2011).
\bibitem{Shirley_65}  J. H. Shirley, Phys. Rev. {\bf 138}, B979 (1965).
\bibitem{Hanggi_rev} M. Grifoni and P. H\"anggi, Phys. Rep. {\bf 304}, 229
   (1998); S. Kohler, J. Lehmann, and P. H\"anggi, Phys. Rep. {\bf 406}, 379
   (2005). 
\bibitem{Oka_cur} T. Oka and H. Aoki, Phys. Rev. B {\bf 79}, 081406(R) (2009);
  {\it ibid.} {\bf 79}, 169901(E) (2009).
\bibitem{Syzranov_08} S. V. Syzranov, M. V. Fistul, and K. B. Efetov,
  Phys. Rev. B {\bf 78}, 045407 (2008).
\bibitem{Calvo_APL} H. L. Calvo, H. M. Pastawski, S. Roche, and L. E. F. Foa
  Torres, Appl. Phys. Lett. {\bf 98}, 232103 (2011).
\bibitem{summation} As Ding {\em et al.},\cite{Berakdar_12} when carrying out
  the summation of Green's functions in $k$-space, we
  replace the Green's functions with all $k$ directions 
by the one with $k$ being along current direction.
\bibitem{examine} By comparing our and their results, 
  we find that the conductance peaks in their results at positive
  (negative) energy are shifted by $0.5\omega_0$ ($-0.5\omega_0$) relative to
  ours. However, there are not no singular point even in their incorrect
  formulae of Green's functions [i.e., Eqs.~(A.3)-(A.8) in
  Ref.~\onlinecite{Berakdar_12}] at the energies related to their peaks. This
  indicates that there must be more errors in their calculations. 
\bibitem{Neto_rev_09} A. H. Castro Neto, F. Guinea, N. M. R. Peres,
  K. S. Novoselov, and A. K. Geim, Rev. Mod. Phys. {\bf 81}, 109 (2009).

\end{thebibliography}
\end{document}